\documentclass [12pt]{article}
\usepackage {graphicx}

\begin{document}
\begin{center}
\textbf{Peculiarities of electronic heat capacity of thulium cuprates}

\textbf{in pseudogap state}
\end{center}

\bigskip

\begin{center}
E.B. Amitin, K.R. Zhdanov, M.Y. Kameneva,

Yu.A. Kovalevskaya, L.P. Kozeeva, I.E. Paukov

\textit{Institute of Inorganic Chemistry, 630090, Novosibirsk, Russia}

A.G. Blinov

\textit{Novosibirsk State University, 630090, Novosibirsk, Russia}

E-mail:\underline {amitin@casper.che.nsk.su}
\end{center}

\bigskip

\paragraph{Abstract}

\bigskip

Precise calorimetric measurements have been carried out in the 7 - 300 K 
temperature range on two ceramic samples of thulium 123 cuprates 
TmBa$_{2}$Cu$_{3}$O$_{6.92}$ and TmBa$_{2}$Cu$_{3}$O$_{6.70}$. The 
temperature dependence of the heat capacity was analyzed in the region where 
the pseudogap state (PGS) takes place. The lattice contribution was 
subtracted from the experimental data. The PGS component has been obtained 
by comparing electronic heat capacities of two investigated samples because 
the PGS contribution for the 6.92 sample is negligible. The anomalous 
behavior of the electronic heat capacity near the temperature boundary of 
PGS was found. It is supposed that this anomaly is due to peculiarities in 
$N(E)$ function where $N$ is the density of electronic states and $E$ is the energy of 
carriers of charge.

\bigskip


\bigskip

\textbf{Introduction}

\bigskip

Last years a number of articles devoted to research of features of various 
properties, connected with so called pseudogap state (PGS), have appeared in 
periodicals. The PGS arises in underdoped yttrium 1-2-3 and lanthanum 2-1-4 
superconductors in the temperature region above the line of superconducting 
transition ($T_{c} < $ $T < $ $T$*) and at concentrations of carriers of charge, 
smaller than the critical value $p_{c}$ appropriate to the point of optimal 
doping (POD) (Fig. 1).

Hundreds publications were devoted to studying of unusual manifestations of 
the PGS in various properties. The most convincing proofs of existence of 
the PGS were received by direct methods - photoemission with the angular 
resolution (ARPES) and methods of electronic tunneling [1-3]. These data 
make evident an appreciable decrease of the density of electronic states in 
the field of PGS. The density of electronic states near the Fermi level has 
V-shaped minimum. If temperature increases and gets over a boundary line 
$T$*$(p)$, this minimum disappears and the system goes out of the PGS. 

At first the line $T$*($p)$ was perceived as a boundary of the region where NMR 
spin-lattice relaxation deviates significantly from the Korringa rule [4, 
5]. Later it has been shown that experimental investigations of different 
physical properties determine different standings of the boundary of this 
state [6]. The border derived from the NMR measurements of the Knight shift 
or from the measurements of spin-lattice relaxation is marked in Fig. 1 with 
$T_{cr}(p)$. It is situated at higher temperatures, than the border derived 
from thermodynamic and kinetic properties $T$*($p)$. Such a discrepancy allows 
several researchers to suggest that there are two regions on the phase 
diagram, corresponding to weak and strong PGS [6]. Discussing the problem, 
we use the notations of ref. [6] throughout the work. The upper temperature 
boundary denoted by $T_{cr}$, is defined as a border of crossover phenomena 
in NMR experiments. The lower boundary is denoted traditionally by 
$T^{\ast }(p)$.

Two probable mechanisms of the PGS formation are discussed in the 
literature:

1. The Cooper pairs exist above the temperature of superconducting phase 
transition forming the incoherent paired states. The number of the pairs 
increases with decreasing temperature and at the temperature $T_{c}$ the 
incoherent pairing transforms into the correlated superconducting state [7, 
8].

2. The charge carriers interact with the fluctuations of magnetic or charge 
short-range order (spin or charge density waves) [9-11]. Such an interaction 
can yield the V-shaped minimum in the density of electron states near the 
Fermi level [12]. 

It is difficult to recognize what mechanism of PGS formation is actual. Many 
authors suppose the second mechanism to be more plausible, than first [13]. 

Along with direct examinations of features of electronic states in PGS of 
1-2-3 and 2-1-4 cuprate HTSC by the methods of ARPES and electronic 
tunneling, the studying of thermodynamic properties of these substances can 
make a considerable contribution to the insight into the subject. The 
investigation of unusual peculiarities in thermodynamic properties of 1-2-3 
and 2-1-4 cuprate HTSC was started by Junod and co-workers in the early 90s 
[14, 15]. Main investigations were carried out by Loram and co-workers after 
the middle of 90s [16-20] and are prolonged now. Such investigations are of 
great interest because of many problems in the theoretical interpretation of 
the phenomena observed. Another problem is the separation of the electronic 
contribution to the heat capacity. For the normal state above $T_{c}$, the 
electronic contribution $C_{el}$ is not greater than 2 - 3 {\%} of total heat 
capacity. Change of $C_{el}$ connected with PGS formation is significantly 
less, namely 0.2 - 0.3 {\%} of total heat capacity. Thus, even minor error 
in the separation of lattice, electronic, and magnetic contributions can 
lead to a significant error in the values describing electronic properties 
of cuprate HTSC.

To determine the electronic contribution at the PGS in yttrium cuprates 
YBCO$_{6 + X}$, Loram and co-workers used the lattice heat capacity 
$C_{lat}$ of dielectric YBCO$_{6}$ as a reference value. Probable differences 
between the values for a reference substance and substance under 
investigation were corrected using the sum of Einstein functions. The 
parameters in the functions were derived from the fitting of experimental 
data at low temperatures. The extrapolation of these components into field 
of PGS has allowed authors to separate electronic heat capacity $C_{el}$ of 
the cuprates investigated. This way of calculation yields the value of 
$C_{el}$ containing also the differences in magnetic and anharmonic 
contributions. The approximation of the difference by Einstein functions and 
its extrapolation in the field of PGS can produce significant errors in the 
electronic contribution calculated.

The aim of this work is to use the experimental results for the sample close 
to the optimal doping as the reference data. We think that this way of 
calculation of electronic contribution in the PGS is more correct.

\bigskip

\textbf{Experimental}

\bigskip

The calorimetric measurements of TmBa$_{2}$Cu$_{3}$O$_{6.92}$ and 
TmBa$_{2}$Cu$_{3}$O$_{6.70}$ were carried out using automatic 
low-temperature vacuum adiabatic calorimetric system. The equipment was 
described elsewhere [21] but only one exception: the calorimeter we used was 
made of copper and covered with silver instead of that made of pure nickel 
[21]. Internal volume of the calorimeter was about 6 cm$^{3}$. Heat capacity 
of empty calorimeter was measured in the temperature range 6 - 310 K. The 
accuracy of the calorimetric results was tested using the standard reference 
sample - benzoic acid. Our results agree with reliable data for benzoic acid 
taken from the literature [22-24] within the limits of $\pm $2 {\%} over the 
temperature range 6 - 10 K, 0.5 {\%} for 10 - 30 K, 0.2 {\%} for 30 - 60 K, 
and 0.1 {\%} above 60 K. During the measurements, temperature increase at 
the heat pulse was 1 to 2 K at $T < $ 30 K and 3 - 5 K over the temperature 
range 30 - 300 K. The average deviation of the experimental values from the 
smoothed curve was about 0.02 {\%} at temperatures 100 -- 300 K. The 
deviation increased with decreasing temperature, growing up to 1 {\%} at 10 
K. A total of 300 - 400 experimental points were performed for each sample.

Ceramic TmBa$_{2}$Cu$_{3}$O$_{X}$ was synthesized in the solid-state 
reaction according to the standard procedure using high-purity oxides 
Tm$_{2}$O$_{3}$, BaO, and CuO. Before the synthesis, the oxides 
Tm$_{2}$O$_{3}$ and CuO were annealed at 750$^{o}$C and 700$^{o}$C, 
respectively, to remove volatile impurities. The synthesis was performed in 
a corundum crucible in the temperature range from 800 to 900$^{o}$C, step by 
step with an interval of 25$^{o}$C, for 25 hours each. Before each step, the 
sample was thoroughly ground in an agate mortar. To receive the sample with 
optimal oxygen content $X$ = 6.92, the heating was performed in an atmosphere 
with excess amount of oxygen. The sample with $X$ = 6.7 was prepared from the 
sample with $X$ = 6.92 which was annealed in air at $T$ = 590$^{o}$C, then 
quenched into liquid nitrogen and annealed at 100$^{o}$C for two days.

Thulium cuprates were chosen for the investigation instead of yttrium 
cuprates because aluminum impurity in thulium cuprate after the synthesis in 
a corundum crucible was several tens lower than in yttrium cuprate. $X$-ray 
powder diffraction showed ceramics TmBa$_{2}$Cu$_{3}$O$_{X}$ to be nearly 
monophase and to contain the impurities of Tm$_{2}$BaCuO$_{5}$ and 
BaCuO$_{2}$. Total amount of these phases was not greater than 2-3 {\%}. The 
analysis showed that the amount of impurities did not change when the oxygen 
content decreased from 6.92 to 6.70. The analysis of structural factors 
sensing to homogeneity of oxygen distribution (orthorhombicity and 
broadening of basal reflections) has shown a high degree of homogeneity both 
in the sample with $X$ = 6.92 and with $X$ = 6.70.

\bigskip

\textbf{Results and Discussion}

\bigskip

Heat capacity $C_{p}$ is the sum of several contributions

\bigskip

$C_{p}=C_{harm}+C_{anh}+C_{el}+C_{magn}$, (1)

\bigskip

\noindent
where $C_{harm}$ is the contribution from harmonic vibrations in a lattice, 
$C_{anh}$ is that from anharmonicity, $C_{el}$ and $C_{magn}$ are the 
electronic and magnetic contributions, respectively. $C_{el}$ contains the 
contribution from the heat capacity of ordinary Fermi particles ($C_{s})$ 
that constitutes the major portion of $C_{el}$ for the sample 1 with X = 
6.92. Besides $C_{s}$, $C_{el}$ in the sample 2 with $X$ = 6.70 contains the 
contribution from the PGS ($C_{PGS})$. In the sample with $X$ = 6.92, a very 
small contribution of $C_{PGS}$ can also be, since the PGS in 1-2-3 cuprates 
disappears at $X \quad  \cong $ 6.97.

The lattice contribution $C_{harm}$ was calculated by fitting the 
experimental data to the sum of Debye and Einstein functions according to 
the procedure described in [25]. The fitting was performed at low 
temperatures. The sum of the electronic, magnetic, and anharmonic terms 
$\gamma _{anh + el + magn}$ = ($C_{anh}+C_{el}+C_{magn})$/T for the 
samples investigated is shown in Fig. 2 as a function of temperature over 
the temperature range 100 - 300 K. We suppose that the anharmonic and 
magnetic contributions to the heat capacity for both samples are nearly 
identical. At least, we state with certainty that the difference between 
those contributions is much greater for YBCO$_{6}$ and YBCO$_{6 + X}$, 
compared by Loram [16-20], than for TmBa$_{2}$Cu$_{3}$O$_{6.7}$ and 
TmBa$_{2}$Cu$_{3}$O$_{6.92}$. The reason is the difference in structure and 
in magnetic order between the compared yttrium cuprates. 

Our experimental data were treated in the following way. After substraction 
of lattice contribution from total heat capacity, the function $f(T)$ was derived:

\bigskip

$f(T)$ = {\{}$C_{anh}(X$=6.92) / $T+\gamma _{el}(X$=6.92) + 
$C_{magn}(X$=6.92)/$T${\}} --

\bigskip

{\{}$C_{anh}(X$=6.70)/ $T+\gamma _{el}(X$=6.70) + $C_{magn}(X$=6.70)/$T$ {\}}. 
(2)

\bigskip

The function $f(T)$ is shown in Fig. 3. If the anharmonic and magnetic 
contributions are nearly the same, the difference is

\bigskip

$f(T)=\Delta \gamma _{s}-\gamma _{PGS}(X$=6.70). (3)

\bigskip

The first term is positive and increases slightly with increasing 
temperature [26]. Function $\Delta \gamma _{s}(T)$ has no extremum. 
Hence, the wavy anomalies in the Fig. 3 are caused by the second term in 
equation 3. The minimum of the $f(T)$ is the result of the maximum of the function 
$\gamma _{PGS}(X$ = 6.70) that exists in the temperature range 200 - 250 K. 
The $f(T)$ tends to zero as temperature increases up to 300 K indicating that the 
PGS transforms into other state.

Moca and Janko [26] have analyzed the experimental data on electronic heat 
capacity measured by Loram and co-workers. The analysis was based on the 
first point of view mentioned above: the incoherent Cooper pairs exist in 
the PGS. Moca and Janko supposed that $\gamma _{el}$ consists of two 
contributions:

\bigskip

$\gamma _{el}(T)=\gamma _{s}(T)+\gamma _{p}(T)$, (4)

\bigskip

\noindent
where $\gamma _{s}(T)$ is the single-particle contribution from Fermi 
particles. This contribution is similar to the first term in Eq. 3. It 
increases slightly and monotonically with increasing temperature. The second 
term in Eq.4, $\gamma _{p}(T)$ is the contribution from the formation of 
the incoherent Cooper pairs. According to Ref. [26], $\gamma _{p}(T)$ 
depends on temperature like ($T$*/$T)^{3}$exp(-2$T$*/$T)$ with a maximum near $T=T$*. 
According to analysis of the data reported by Loram et al., the authors [26] 
suppose the position of maximum of $\gamma _{el}(T)$ to shift to lower 
temperatures when the oxygen content increases. For the cuprate with $X$ = 6.7, 
the maximum of function $\gamma _{el}(T)$ was estimated to be within the 
temperature range 120 to 130 K [26]. The experimental results in Fig. 3 
clearly show that there is no maximum over the temperature range 120 - 130 
K.

If there is a systematic error in the separation of harmonic and anharmonic 
contributions, the error in the electronic contribution can be as high as 
tens percent. Such an error in the calculations is equivalent to the 
experimental error of 0.1 - 0.2 {\%} in the total heat capacity. It means 
that we analyze the function $C_{p}(T)$ in trying to recognize its 
peculiarities comparable with an accuracy of the experiments. This 
illustrates how complicate is to extract the electronic heat capacity and 
correctly determine its features. 

It is possible to assume, that observed anomalies are related to impurities 
in samples. In fact, copper oxide is a starting material for the synthesis 
and it undergoes two phase transitions, at $T$ = 212.6 K and $T$ = 229.5 K [28]. 
Nevertheless, to produce such an effect in heat capacity, the samples have 
to contain at least 4 or 5 {\%} of copper oxide. We carefully investigated 
the samples by means of X-ray powder diffraction with a limit of detection 
about 2 - 3 {\%} in searching for the copper oxide impurity. No traces of 
CuO were found. Besides, the effect observed in Fig. 3 is in the difference 
between heat capacities of two samples. The sample with $X$ = 6.70 was received 
after gentle heat treatment (at 590$^{o}$C) of the sample with $X$ = 6.92. 
Copper oxide starts to decompose at 1061$^{o}$C and it means that both of 
the samples contain the same amount of CuO. The probable effect of the phase 
transitions in copper oxide should be eliminated in the evaluation of the 
$f(T)$ function (Eq. 2). Thus, the copper oxide impurity cannot explain the 
anomalous behavior of the heat capacity. The amplitude of the anomalous heat 
capacity, by our data, decreases as the sample composition approaches the 
point of optimal doping. These facts lead us to the conclusion that the 
peculiarities observed are connected with the transformation of the samples 
into the PGS.

Recent theoretical investigations examined the electronic systems 
interacting with charge or spin short-range fluctuations [13, 29]. The 
interaction is shown to be responsible for the formation of pseudogap 
structure near the Fermi level. The evaluations based on the mean field 
approximation fail to describe correctly the properties of the system. At 
proper consideration of the fluctuations in the PGS it is found that, 
besides V-shaped minimum, the sharp peaks in the density of electronic 
states can form [13, 29]. It would appear reasonable that the effective 
value $N(E)$ near the Fermi level can be very complex and sensitive to the 
changes in the system near the PGS. In that case the transformation into the 
PGS can go with complex anomalous behavior of various physical properties. 
First of all, these are the properties connected with the density of states 
of charge carriers. We suppose that the observed anomalous behavior of the 
electronic heat capacity is the result of the transformation. Probably, the 
same phenomenon is responsible for unusual function $C_{p}(T)$ for lanthanum 
2-1-4 cuprates described in [27]. We assume that our experimental data for 
thulium cuprates, together with the results on the heat capacity of 
lanthanum 2-1-4 cuprates [27], can be qualitative confirmation of the 
theoretical calculations of Sadovskiy et al.

The work was supported in part by the Russian Foundation for Basic 
Researches (grant N.00-02-17914), by Scientific Programs "High-temperature 
Superconductivity" (grant 98009) and "Universities of Russia" (grant 1785), 
and by the FAP "Integration" (grant 274).

\newpage

\bigskip

\subsubsection*{REFERENCES}

\bigskip

1. Randeria M, Campuzano J.C. cond-mat/9709107.

2. Randeria M, cond-mat/9710223.

3. Ding H. et al. Nature V. \textbf{382}, 51 (1996).

4. Millis A., Monien H., Pines D. Phys.Rev. B, \textbf{42}, 1671 (1990).

5. Gorny K. et al., Phys.Rev.Lett., \textbf{85}, 177 (1999).

6. Schmalian J., Pines D., Stojkovich B., Phys.Rev.B, \textbf{60}, 667 (1999).

7. Geshkenbein V.B., Ioffe L.B., Larkin A.I., Phys.Rev.B, \textbf{55}, 3173 
(1997).

8. Gusynin V.P., Loktev V.M., Sharapov S.G., GETP, \textbf{115}, 1243 (1999).

9. Schmalian J., Pines D., Stojkovich B., Phys.Rev.Lett., \textbf{80}, 3839 
(1998).

10. Kuchinskiy E.Z., Sadovskiy M.V., GETP., \textbf{115}, 1765 (1999); E-prints 
archive, cond-mat/9808321 (1998).

11. Posazgennikova A.I., Sadovskiy M.V., GETP, \textbf{115}, 632 (1999); 
E-prints archive, cond-mat/9806199 (1998).

12. Sadovskiy M., Kuchinskiy E.Z., Physica C, \textbf{341}-\textbf{348}, 879 
(2000).

13. Sadovskiy M.V., Uspehi Phys.Nauk, \textbf{171}, 539 (2001).

14. Junod A., Graf T., Sanchez, et al. Physica B, \textbf{165}-\textbf{166}, 
1335 (1990).

15. Junod A., et al., Physica C, \textbf{168}, 47 (1990).

16. Liang W.Y., Loram J.W., Mirza K.A., et al., Physica C, \textbf{263}, 277 
(1996).

17. Loram J.W., Mirza K.A., Cooper J.R., et al., Physica C, 
\textbf{282}-\textbf{287}, 1405 (1997)

18. Loram J.W.,et al., J. Supercond. \textbf{7}, 234 (1994).

19. Loram J.W., Tallon J.L., Physica C, \textbf{349}, 53 (2001); 
cond-mat./0005063.

20. Loram J.W., Tallon J.L., Williams G.V.M., Physica C, \textbf{338}, 9 (2000).

21. Bessergenev V.G., Kovalevskaya Ju. A., Paukov I. E., Starikov M. I., 
Opperman N., Reichelt W., J. Chem.Therm., \textbf{24}, 85 (1992).

22. Rybkin N.P., Orlova M.P, Baraniuk A.K. Izmeritelnaya tekhnika, No 7, 29 
(1974).

23. Moriya K, Matsuo T, Suga H., J. Chem. Therm. \textbf{14}, 1143 (1982).

24. Sorai M., Kayi K., KanekoY., J.Chem. Therm., \textbf{24}, 167 (1992).

25. Zhdanov K.R., Rahmenkulov F.S., Fedorov V,E., Mischenko A.V. Phyzika 
Tverdogo Tela, \textbf{30}, 1119 (1988).

26. Moca C.P., Janko B., E-prints arXiv: cond-mat/0105202 v1.

27. Loram J.W., et al., J. Phys. Chem. Solids, \textbf{59}, 2091 (1998).

28. Junod A., et al., J. Phys.: Condens. Mat., \textbf{1}, 8021 (1989).

29. Sadovskiy M.V., Kuchinskiy,GETP , \textbf{117}, 613 (2000).

\newpage 
\begin{center}
\textbf{Figures}
\end{center}

\begin{center}
\begin{figure}[htbp]
\includegraphics*[bbllx=0.19in,bblly=0.18in,bburx=4.77in,bbury=4.75in,scale=1.00]{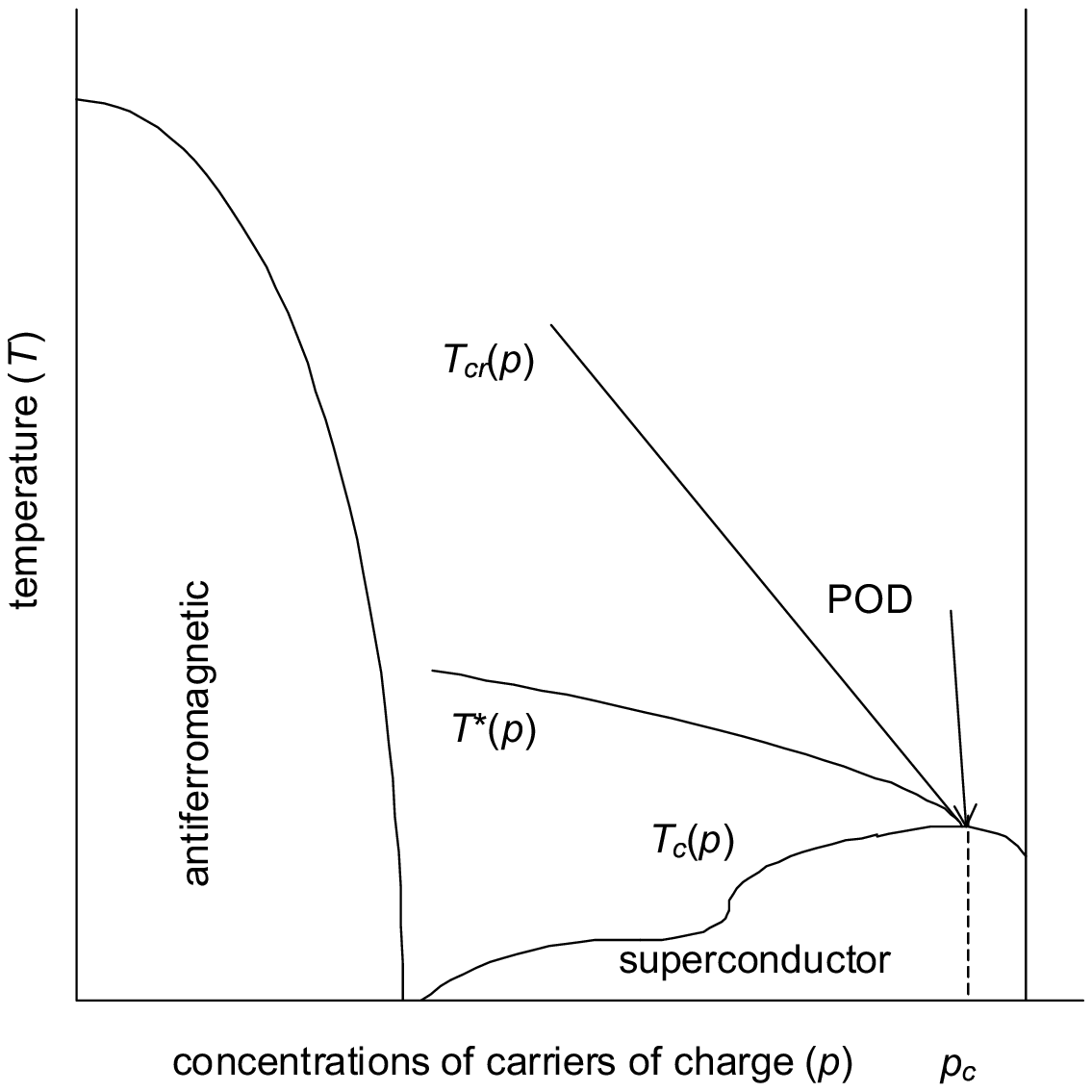}
\label{fig1}
\end{figure}

\end{center}

\bigskip

\begin{center}
Fig. 1. Schematic phase diagram for the cuprates
\end{center}

\newpage 
\begin{center}
\begin{figure}[htbp]
\includegraphics*[bbllx=0.19in,bblly=0.18in,bburx=5.36in,bbury=4.31in,scale=1.00]{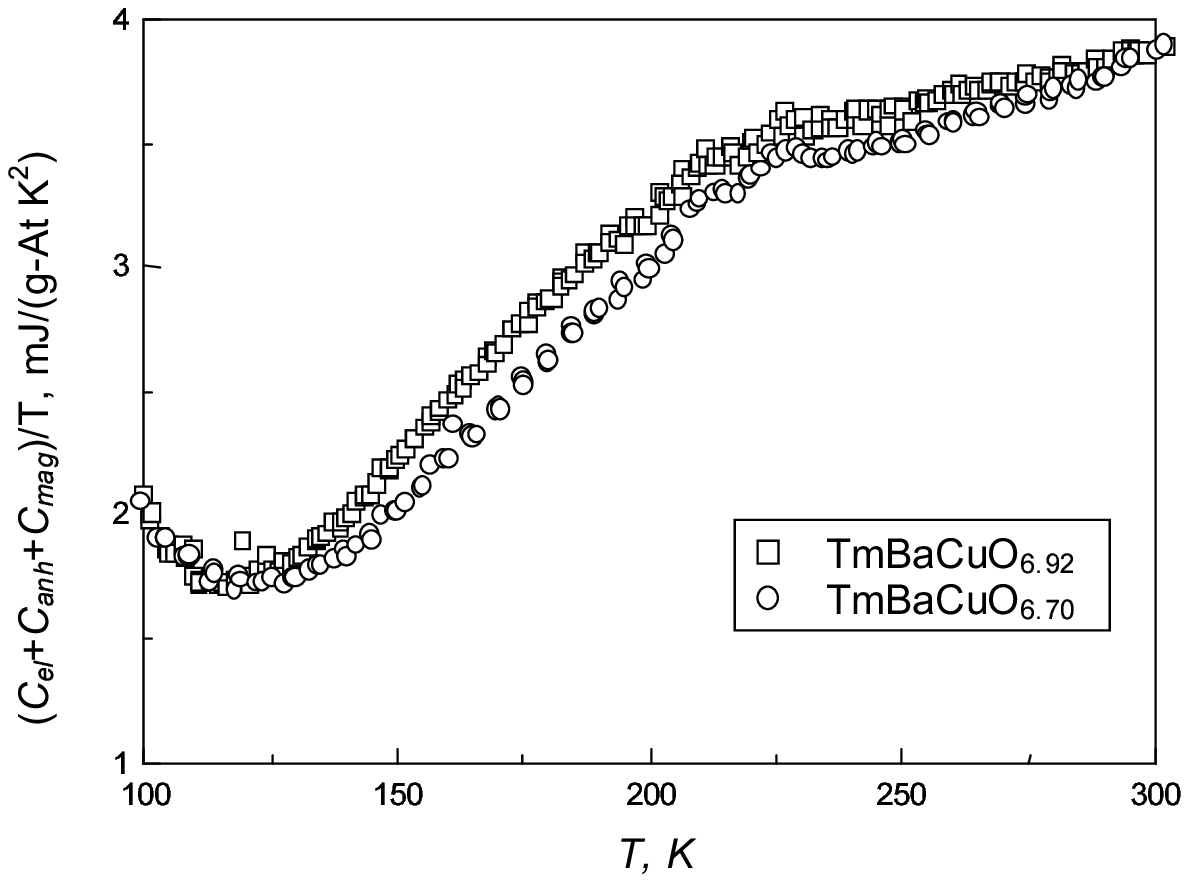}
\label{fig2}
\end{figure}

\end{center}

\bigskip

\begin{center}
Fig. 2. Electronic, anharmonic, and magnetic contributions to the heat 
capacity of thulium cuprates
\end{center}

\newpage 
\begin{center}
\begin{figure}[htbp]
\includegraphics*[bbllx=0.19in,bblly=0.18in,bburx=5.59in,bbury=4.41in,scale=1.00]{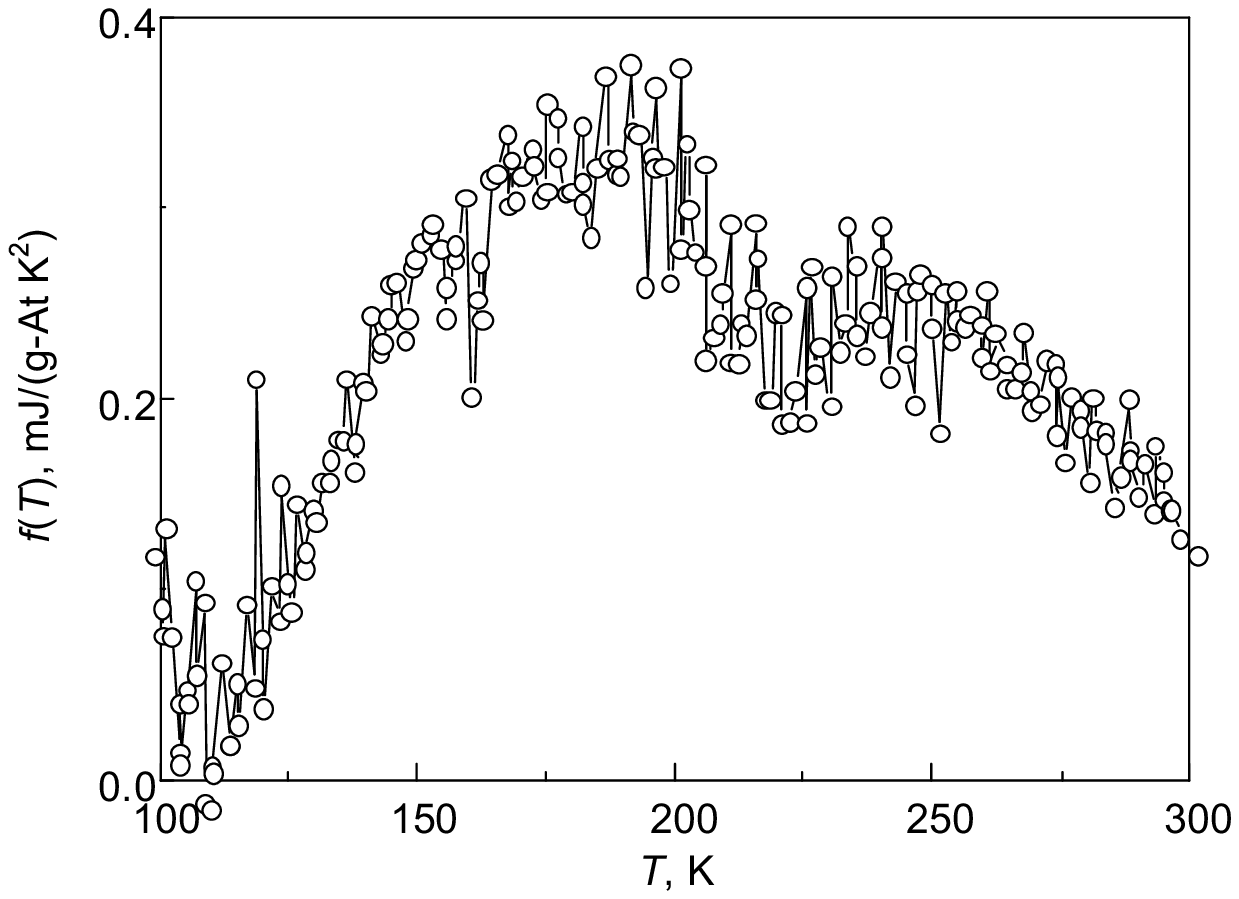}
\label{fig3}
\end{figure}

\end{center}

\bigskip

\begin{center}
Fig. 3. Function $f(T)$
\end{center}

\end{document}